\documentstyle[preprint,aps,epsfig]{revtex}
\tighten
\begin{document}
%__________________________________________________
%\twocolumn
%_______________________ Title, Authors ____________________________________
 \preprint{Preprint Numbers: \parbox[t]{45mm}{ANL-PHY-8864-TH-97\\
                                             MPG-VT-UR 120/97\\
                                           % nucl-th/9801059
}}

\title{Chemical potential dependence of $\pi$ and $\rho$ properties}

\author{P. Maris\footnotemark[1], 
C. D. Roberts\footnotemark[1] and 
S. Schmidt\footnotemark[2]\vspace*{0.2\baselineskip}} 
\address{\footnotemark[1]Physics Division, Bldg. 203, Argonne National
Laboratory, Argonne IL 60439-4843\\\vspace*{0.2\baselineskip} 
\footnotemark[2]Fachbereich Physik, Universit\"at Rostock, D-18051 Rostock,
Germany\vspace*{0.2\baselineskip} }
\date{10th December 1997}
\maketitle
%-------------------------------------------------------------------
\begin{abstract}
Using a confining, Dyson-Schwinger equation model of QCD at finite
temperature $(T)$ and chemical potential $(\mu)$ we study the behaviour of
$\langle \bar q q \rangle$, $m_\pi$, $f_\pi$ and the masses of the $\rho$-
and $\phi$-mesons.  For each of these quantities there is a necessary
anticorrelation between its response to $T$ and its response to $\mu$.
$(-\langle \bar q q \rangle)$ and $f_\pi$ decrease with $T$ and increase with
$\mu$; $m_\pi$ is almost insensitive to $T$ and $\mu$ until very near the
border of the confinement domain; and the mass of the longitudinal component
of the vector mesons increases with $T$ and decreases with $\mu$.  At $T=0$,
the $\rho$-meson mass is reduced by approximately 15\% at nuclear matter
density.  These results are a consequence of the necessary
momentum-dependence of the dressed-quark self-energy.
\end{abstract}
%..............................
%\begin{tabular}[h]{ccc}
%\hspace*{-3mm}\parbox{87mm}{
%
High-energy heavy-ion ($A$-$A$) experiments, which produce systems with large
baryon density, are an important preliminary step in the search for a
quark-gluon plasma.  An outcome of these experiments is the observation that
the dilepton yield in the region below the $\rho$-resonance is approximately
five-times greater than that seen in proton induced ($p$-$A$)
reactions~\cite{drees97}.  One model calculation~\cite{brown95} shows that
this enhancement can be explained by a medium-induced reduction of the
$\rho$-meson's mass, another~\cite{rapp97} that it follows from an increase
in the $\rho$-meson's width.  A decrease in the $\rho$-meson's mass is
consistent with the QCD sum rules analysis of Ref.~\cite{derek95} but
inconsistent with that of Ref.~\cite{klingl97}, which employs a more complex
phenomenological model for the in-medium spectral density used in matching
the two sides of the sum rule.  In Ref.~\cite{klingl97} there is no shift in
the $\rho$-meson mass but a significant increase in its width.  The
consistency between Refs.~\cite{rapp97} and \cite{klingl97} is not surprising
since, in contrast to Ref.~\cite{derek95}, they both rely heavily on
effective Lagrangians with elementary hadron degrees-of-freedom.  Herein,
using a simple, Dyson-Schwinger equation (DSE) model of QCD, we reconsider
the response of the $\rho$-meson's mass to increasing baryon density.  In
focusing on dressed-quark and -gluon degrees of freedom, our study has
similarities to that of Ref.~\cite{derek95}.

DSEs provide a nonperturbative, continuum framework for analysing quantum
field theories.  While they have been used extensively at $T=0=\mu$ in the
study of dynamical chiral symmetry breaking (DCSB) and
confinement~\cite{dserev}, and the calculation of hadronic
observables~\cite{pctrev}, their application at finite $T$ and $\mu$, though
straightforward, is in its infancy.

We begin with the specification of a model dressed-gluon propagator
\begin{equation}
\label{mnprop}
g^2 D_{\mu\nu}(\vec{p},\Omega_k) = 
\left(\delta_{\mu\nu} 
- \frac{p_\mu p_\nu}{|\vec{p}|^2+ \Omega_k^2} \right)
2 \pi^3 \,\frac{\eta^2}{T}\, \delta_{k0}\, \delta^3(\vec{p})\,,
\end{equation}
where $(p_\mu):=(\vec{p},\Omega_k)$, $\Omega_k = 2 k \pi T$ is the boson
Matsubara frequency and $\eta$ is a mass-scale parameter.  
%.....
% } & \hspace*{3mm}&
%
% \parbox{87mm}{
%.....
Equation (\ref{mnprop}) is an extension to finite-$T$ of the model for
$D_{\mu\nu}$ introduced in Ref.~\cite{mn83}.  It has the feature that the
infrared enhancement of the dressed-gluon propagator suggested by the DSE
studies of Refs.~\cite{gpropir} is manifest.  However, it represents poorly
the behaviour of $D_{\mu\nu}(\vec{p},\Omega_k)$ away from $|\vec{p}|^2+
\Omega_k^2 \approx 0$.  Nevertheless, this limitation leads only to easily
identifiable artefacts and hence does not preclude the judicious use of the
model.  Another feature of (\ref{mnprop}) is that it specifies a model with
the merit of simplicity: it provides for the reduction of integral DSEs
(e.g., the gap equation and Bethe-Salpeter equations) to algebraic equations,
which facilitates the elucidation of many of the qualitative features of more
sophisticated models.

Using (\ref{mnprop}) and the rainbow-truncation for the dressed-quark-gluon
vertex: $\Gamma_\mu^a(k,p)= \case{1}{2}\lambda^a \gamma_\mu$, the DSE for the
dressed-quark propagator in our model for QCD at finite-$T$ and $\mu$
(QCD$_T^\mu$) is~\cite{brs96}
\begin{equation}
\label{mndse}
S^{-1}(\vec{p},\omega_k) = S_0^{-1}(\vec{p},\omega_k)
        + \case{1}{4}\eta^2\gamma_\nu S(\vec{p},\omega_k) \gamma_\nu\,,
\end{equation}
where $S_0^{-1}(\vec{p},\omega_k):= i\vec{\gamma}\cdot\vec{p}+i\gamma_4
\omega_{k+}+m$, with $\omega_{k+}= \omega_k + i \mu$, $\omega_{k} = (2k +
1)\pi T$ is the fermion Matsubara frequency and $\mu$ is the chemical
potential, and $m$ is the current-quark mass.  (In our Euclidean formulation,
$\{\gamma_\mu,\gamma_\nu\}=2\delta_{\mu\nu}$ with $\gamma_\mu^\dagger =
\gamma_\mu$.)  The solution of (\ref{mndse}) has the general form
\begin{eqnarray}
\label{qprop}
S(\tilde p_k) & = & \frac{1}{i\vec{\gamma}\cdot\vec{p}\, A(\tilde p_k) 
+ i\gamma_4 \omega_{k+}\, C(\tilde p_k) 
+ B(\tilde p_k) }\,,\\
& = & -i \vec{\gamma}\cdot\vec{p} \, \sigma_A(\tilde p_k) 
- i\gamma_4 \omega_{k+}\, \sigma_C(\tilde p_k) 
+ \sigma_B(\tilde p_k)\,,
\end{eqnarray}
where $\tilde p_k := (\vec{p},\omega_{k+})$, and (\ref{mndse}) entails that
the scalar functions introduced in (\ref{qprop}) satisfy
\begin{eqnarray}
\label{beqnfour}
\eta^2 m^2 & = & B^4 + m B^3 + \left(4 \tilde p_k^2 - \eta^2 -
        m^2\right) B^2 -m\,\left( 2\,{{\eta }^2} + {m^2} +
        4\,\tilde p_k^2 \right)B   \,,     \\ 
\label{aeqc}
A(\tilde p_k) & = & C(\tilde p_k) = \frac{2 B(\tilde p_k)}{m +B(\tilde p_k)}\,.
\end{eqnarray}

In the chiral limit $(m=0)$ the character of the solution   
%
% }\end{tabular}
%
% \twocolumn
%
\hspace*{-\parskip}of (\ref{mndse}) is transparent.  There are two
qualitatively distinct solutions.  The ``Nambu-Goldstone'' solution, for
which
\begin{eqnarray}
\label{ngsoln}
B_0(\tilde p_k) & = &\left\{
\begin{array}{lcl}
\sqrt{\eta^2 - 4 \tilde p_k^2}\,, & &{\sf Re}(\tilde p_k^2)<\case{\eta^2}{4}\\
0\,, & & {\rm otherwise}
\end{array}\right.\\
\label{cng}
C_0(\tilde p_k) & = &\left\{
\begin{array}{lcl}
2\,, & & {\sf Re}(\tilde p_k^2)<\case{\eta^2}{4}\\
\case{1}{2}\left( 1 + \sqrt{1 + \case{2 \eta^2}{\tilde p_k^2}}\right)
\,,& & {\rm otherwise}\,,
\end{array}\right.
\end{eqnarray}
describes a phase of this model in which: 1) chiral symmetry is dynamically
broken, because one has a nonzero quark mass-function, $B(\tilde p_k)$, in
the absence of a current-quark mass; and 2) the dressed-quarks are confined,
because the propagator prescribed by these functions does not have a Lehmann
representation.  The alternative ``Wigner'' solution, for which
\begin{eqnarray}
\label{wsoln}
\hat B_0(\tilde p_k)  \equiv  0& , \; & 
\hat C_0(\tilde p_k)  = 
\case{1}{2}\left( 1 + \sqrt{1 + \case{2 \eta^2}{\tilde p_k^2}}\right)\,,
\end{eqnarray}
describes a phase of the model in which chiral symmetry is not broken and the
dressed-quarks are not confined.

With these two phases, characterised by qualitatively different,
momentum-dependent modifications of the quark propagator, this model of
QCD$_T^\mu$ can be used to explore simultaneously both chiral symmetry
restoration and deconfinement.  That and the calculation of its equilibrium
thermodynamic properties are the subject of Ref.~\cite{blaschke97}.  The
model exhibits coincident, first order deconfinement and chiral symmetry
restoration for all $\mu\neq 0$ but the coincident transitions are second
order for $\mu=0$.  The extreme points on the phase boundary are $(T=0,\mu
\approx 0.28\, \eta)$ and $(T\approx 0.16 \,\eta, \mu=0)$.

The vacuum quark condensate is proportional to the matrix trace of the
chiral-limit dressed-quark propagator.  Using (\ref{qprop}), (\ref{ngsoln})
and (\ref{cng}) we obtain the following expression, valid in the domain of
confinement and DCSB
\begin{equation}
\label{qbq}
-\langle \bar q q \rangle = 
\eta^3\,\frac{8 N_c}{\pi^2} \bar T\,\sum_{l=0}^{l_{\rm max}}\,
\int_0^{\bar\Lambda_l}\,dy\, y^2\,
{\sf Re}\left( \sqrt{\case{1}{4}- y^2 - \bar\omega_{l+}^2 }\right)\,,
\end{equation}
with: $\bar T=T/\eta$, $\bar \mu=\mu/\eta$, $\bar\omega_{l+}=
\omega_{l+}/\eta$; and $\bar\omega^2_{l_{\rm max}}\leq
\case{1}{4}+\bar\mu^2$, $\bar\Lambda_l^2 =
\case{1}{4}+\bar\mu^2-\bar\omega_l^2$.
For $T=0=\mu$, $(-\langle \bar q q \rangle) = \eta^3 /(80\,\pi^2) = (0.11\,
\eta)^3$.
In Fig.~\ref{condensate} we see that $(-\langle \bar q q \rangle)$ decreases
with $T$ but {\it increases} with increasing $\mu$, up to a critical value of
$\mu_c(T)$ when it drops discontinuously to zero.\footnote{These results are
in qualitative and semiquantitative agreement with $(T= 0,\mu\neq
0)$~\protect\cite{greg97} and $(T\neq 0,\mu = 0)$~\protect\cite{bender96}
studies of a more sophisticated model that better represents the behaviour of
$D_{\mu\nu}$ in the ultraviolet.  The $\mu$-dependence is also qualitatively
identical to that observed in a random matrix theory with the global
symmetries of the QCD partition function~\protect\cite{jackson}.}  The
increase of $(-\langle \bar q q \rangle)$ with $\mu$ must be expected in the
confinement domain because confinement entails that each additional quark
must be locally-paired with an antiquark, thereby increasing the density of
condensate pairs.  This vacuum rearrangement is manifest in the behaviour of
the necessarily-momentum-dependent scalar part of the quark self energy,
$B(\tilde p_k)$.

Our primary interests are the bound state properties of $\pi$- and
$\rho$-mesons.  The $\pi$ has been much studied and it follows~\cite{maris97}
from the axial-vector Ward-Takahashi identity that, writing the $\pi$
Bethe-Salpeter amplitude as
\begin{equation}
\label{approxgamma}
\Gamma_\pi(p;P) = i\gamma_5\,B_0(p^2)\,,
\end{equation}
one can obtain a reliable approximation in the calculation of those $\pi$
observables for which the dominant, intrinsic momentum-scale is less-than
10$\,$GeV$^2$.  Using (\ref{approxgamma}) then~\cite{greg97,bender96} in the
domain of confinement and DCSB
\begin{eqnarray}
\label{pimasslhs}
f_\pi^2 m_\pi^2 & = &\langle m\,\bar q q \rangle_\pi\,; \\
\langle m\,\bar q q\rangle_\pi
& =&
\label{pimass}
\eta^4\,\frac{8 N_c }{\pi^2} 
\bar T\,\sum_{l=0}^{l_{\rm max}}\,
\int_0^{\bar\Lambda_l}\,dy\, y^2\,{\sf Re} \left\{\bar B_0
\left(\bar\sigma_{B_0} 
        - \bar B_0 \left[ \bar\omega_{l+}^2 \bar\sigma_C^2 
        + y^2 \bar\sigma_A^2 + \bar\sigma_B^2\right]\right)\right\}\,,
\end{eqnarray}
where $|\vec{p}|= \eta \,y$ and $B_0(\tilde p_l):= \eta\,\bar
B_0(\eta\,\vec{y},\eta\,\bar \omega_l)$, etc.  The right-hand-side of
(\ref{pimass}) is zero for $m=0$ and increases linearly with $m$, for
small-$m$.  In (\ref{pimasslhs}) the canonical normalisation constant for the
$\pi$ Bethe-Salpeter amplitude is
\begin{eqnarray}
\label{npisq}
\lefteqn{f_\pi^2  =  
\eta^2\,\frac{2 N_c }{\pi^2} 
\bar T\,\sum_{l=0}^{l_{\rm max}}\,
\int_0^{\bar\Lambda_l}\,dy\, y^2\,
{\sf Re} \left[
\bar B_0^2\,
\left\{\bar\sigma_A^2 - 2 \left[ 
        \bar\omega_{l+}^2\bar\sigma_C\bar\sigma_C^\prime + 
 y^2 \bar\sigma_A\bar\sigma_A^\prime + \bar\sigma_B\bar\sigma_B^\prime
        \right] \right.\right.} \\
& & \nonumber \left.\left.\!
   - \case{4}{3}\,y^2\,\left(
        \left[\bar \omega_{l+}^2\left(\bar\sigma_C\bar\sigma_C^{\prime\prime} -
        (\bar\sigma_C^\prime)^2\right) + 
        y^2\left(\bar\sigma_A\bar\sigma_A^{\prime\prime} -
        (\bar\sigma_A^\prime)^2\right) + 
        \bar\sigma_B\bar\sigma_B^{\prime\prime} -
        (\bar\sigma_B^\prime)^2 \right] \right) \right\}\right],
\end{eqnarray}
with $\bar\sigma^\prime:= \partial \sigma(y,\bar\omega_{l+})/\partial y$,
which provides a quantitatively accurate approximation to the leptonic decay
constant.\footnote{The relation between the normalisation of the $\pi$
Bethe-Salpeter amplitude and the leptonic decay constant is discussed in
Ref.~\protect\cite{maris97}.  The demonstration\protect\cite{cdrunpub} that
(\protect\ref{npisq}) provides an accurate estimate of the $\pi$ decay
constant when using (\protect\ref{approxgamma}) is an antecedent to
Ref.~\protect\cite{maris97}.}

In the chiral limit, we have from (\ref{aeqc})-(\ref{cng}) that
\begin{equation}
\label{chiralident}
\bar\sigma_{B_0} = B_0 ,\;
        \bar\sigma^\prime_{B_0}= - \frac{2}{B_0},\;
        \bar\sigma^{\prime\prime}_{B_0} = - \frac{4}{B_0^3}\,,
\end{equation}
$\bar\sigma_C = C = \bar\sigma_A$ and
$\bar\sigma_C^{\prime}=0=\bar\sigma_A^{\prime}$.  The simplicity of the model
is again manifest in these identities, which yield
\begin{eqnarray}
\label{npialg}
f_\pi^2 & = & \eta^2 \frac{16 N_c }{\pi^2} 
\bar T\,\sum_{l=0}^{l_{\rm max}}\,
\frac{\bar\Lambda_l^3}{3}
\left( 1 + 4 \,\bar\mu^2 - 4 \,\bar\omega_l^2 - \case{8}{5}\,\bar\Lambda_l^2
\right)\,.
\end{eqnarray}
Characteristic in (\ref{npialg}) is the combination $\mu^2 - \omega_l^2$,
which entails that, whatever change $f_\pi$ undergoes as $T$ is increased,
the {\it opposite} occurs as $\mu$ is increased.  Without calculation,
(\ref{npialg}) indicates that $f_\pi$ will {\it decrease} with $T$ and {\it
increase} with $\mu$, and this provides a simple elucidation of the
calculated results in Refs.~\protect\cite{greg97,bender96}.
Figure~\ref{fpimpi} illustrates this behaviour for $m\neq 0$.

In Fig.~\ref{fpimpi} we also plot $m_\pi$, from (\ref{pimasslhs}).  It is
{\it insensitive} to changes in $\mu$ and only increases slowly with $T$,
until $T$ is very near the critical temperature.  This insensitivity is the
result of mutually cancelling increases in $\langle m\,\bar q q\rangle_\pi$
and $f_\pi$, and is a feature of studies that preserve the
momentum-dependence of the confined, dressed-quark degrees of freedom in
meson bound states.

With $\eta = 1.37\,$GeV and $m=30\,$MeV, one obtains $f_\pi=92\,$MeV and
$m_\pi= 140\,$MeV at $T=0=\mu$.  That large values of $\eta$ and $m$ are
required is a quantitative consequence of the inadequacy of
(\protect\ref{mnprop}) in the ultraviolet: the large-$p^2$ behaviour of the
scalar part of the dressed-quark self-energy is incorrect.  This defect is
remedied easily~\cite{maris97} without qualitative changes to the results
presented here~\cite{maris98}.

With the vector Ward identity unable to assist with a significant
simplification of the bound state problem, $\rho$-meson properties are more
difficult to study: one must solve directly the vector-meson Bethe-Salpeter
equation.  The ladder truncation of the kernel in the inhomogeneous
axial-vector vertex equation and the rainbow truncation of the quark DSE form
an axial-vector Ward-Takahashi identity preserving pair~\cite{brs96}.  It
follows that the ladder BSE is accurate for flavour-nonsinglet pseudoscalar
and vector bound states of equal-mass quarks because of a cancellation in
these channels between diagrams of higher order in the skeleton expansion of
which this pair of truncations is the lowest order term.

A ladder BSE using the $T=0$ limit of (\ref{mnprop}) was introduced in
Ref.~\cite{mn83}.  It has one notable pathology: the bound state mass is
determined only upon the additional specification that the constituents have
zero relative momentum.  This specification leads to a conflict with the
axial-vector Ward-Takahashi identity, which relates the momentum dependence
of the $\pi$ Bethe-Salpeter amplitude to the functions in the quark
propagator~\cite{maris97}.  We find this to be an artefact of implementing
the delta-function limit discontinuously; i.e., the identities~\cite{maris97}
between the scalar functions in the $\pi$ Bethe-Salpeter amplitude, and
$A(p^2)$ and $B(p^2)$ are manifest for any finite-width representation of the
delta-function, as this width is reduced continuously to zero.  In other
respects this ladder BSE provides a useful qualitative and semi-quantitative
tool for analysing features of the pseudoscalar and vector meson masses.  For
example, Goldstone's theorem is manifest, in that the $\pi$ is massless in
the chiral limit, and also $m_\pi^2$ rises linearly with the current-quark
mass.  Further, there is a naturally large splitting between $m_\pi$ and
$m_\rho$, which decreases slowly with the current-quark mass.

To illustrate this and determine the response of $m_\rho$ to increasing $T$
and $\mu$, we generalise the BSE of Ref.~\cite{mn83} to finite-$(T,\mu)$ as
\begin{equation}
\label{bse}
\Gamma_M(\tilde p_k;\check P_\ell)= - \frac{\eta^2}{4}\,
{\sf Re}\left\{\gamma_\mu\,
S(\tilde p_i +\case{1}{2} \check P_\ell)\,
\Gamma_M(\tilde p_i;\check P_\ell)\,
S(\tilde p_i -\case{1}{2} \check P_\ell)\,\gamma_\mu\right\}\,,
\end{equation}
where $\check P_\ell := (\vec{P},\Omega_\ell)$.  The bound state mass is
obtained by considering $\check P_{\ell=0}$ and, in ladder truncation, the
$\rho$- and $\omega$-mesons are degenerate.

As a test we first consider the $\pi$ equation, which admits the solution
\begin{equation} 
\Gamma_\pi(P_0) = \gamma_5 \left(i \theta_1 
        + \vec{\gamma}\cdot \vec{P} \,\theta_2 \right)
\end{equation}
and yields the mass plotted in Fig.~\ref{pirhomass}.  The mass behaves in
qualitatively the same manner as $m_\pi$ in Fig.~\ref{fpimpi}, from
(\ref{pimasslhs}), as required if (\ref{bse}) is to provide a reliable guide.
In particular, it vanishes in the chiral limit.

In the case of the $\rho$-meson there are two components: longitudinal and
transverse to $\vec{P}$.  The BSE has a solution of the form
\begin{equation}
\Gamma_\rho = \left\{
\begin{array}{l}
\gamma_4 \,\theta_{\rho+} \\
\left(
\vec{\gamma} - \case{1}{|\vec{P}|^2}\,\vec{P} \vec{\gamma}\cdot\vec{P}\right)\,
        \theta_{\rho-}
\end{array}
\right.\,,
\end{equation}
where $\theta_{\rho+}$ labels the longitudinal and $\theta_{\rho-}$ the
transverse solution.  The eigenvalue equation obtained from (\ref{bse}) for
the bound state mass, $M_{\rho\pm}$, is
\begin{equation}
\label{rhomass}
\frac{\eta^2}{2}\,{\sf Re}\left\{ \sigma_S(\omega_{0+}^2
        - \case{1}{4} M_{\rho\pm}^2)^2 
- \left[ \pm \,\omega_{0+}^2 - \case{1}{4} M_{\rho\pm}^2\right]
        \sigma_V(\omega_{0+}^2- \case{1}{4} M_{\rho\pm}^2)^2 \right\}
= 1\,.
\end{equation}

The equation for the transverse component is obtained with $[- \omega_{0+}^2
- \case{1}{4} M_{\rho-}^2]$ in (\ref{rhomass}).  Using the chiral-limit
identities, (\ref{chiralident}), one obtains immediately that
\begin{equation}
M_{\rho-}^2 = \case{1}{2}\,\eta^2,\;\mbox{{\it independent} of $T$ and $\mu$.}
\end{equation}
This is the $T=0=\mu$ result of Ref.~\cite{mn83}.  Even for nonzero
current-quark mass, $M_{\rho-}$ changes by less than 1\% as $T$ and $\mu$ are
increased from zero toward their critical values.  Its insensitivity is
consistent with the absence of a constant mass-shift in the transverse
polarisation tensor for a gauge-boson.

For the longitudinal component one obtains in the chiral limit:
\begin{equation}
\label{mplus}
M_{\rho+}^2 = \case{1}{2} \eta^2 - 4 (\mu^2 - \pi^2 T^2)\,.
\end{equation}
The characteristic combination $[\mu^2 - \pi^2 T^2]$ again indicates the
anticorrelation between the response of $M_{\rho+}$ to $T$ and its response
to $\mu$, and, like a gauge-boson Debye mass, that $M_{\rho+}^2$ rises
linearly with $T^2$ for $\mu=0$.  The $m\neq 0$ solution of (\ref{rhomass})
for the longitudinal component is plotted in Fig.~\ref{pirhomass}.  As
signalled by (\ref{mplus}), $M_{\rho+}$ {\it increases} with increasing $T$
and {\it decreases} as $\mu$ increases.\footnote{There is a 25\% difference
between the value of $\eta$ required to obtain the $T=0=\mu$ values of
$m_\pi$ and $f_\pi$, from (\protect\ref{pimass}) and (\protect\ref{npisq}),
and that required to give $M_{\rho\pm}= 0.77\,$GeV.  This is a measure of the
quantitative accuracy of our algebraic model.}

We have stated that contributions from skeleton diagrams not included in the
ladder truncation of the vector meson BSE do not alter the calculated mass
significantly because of cancellations between these higher order
terms~\cite{brs96}.  This is illustrated explicitly in two calculations:
Ref.~\cite{mitchell97}, which shows that the $\rho\to\pi\pi\to\rho$
contribution to the real part of the $\rho$ self-energy; i.e., the
$\pi$-$\pi$ induced mass-shift, is only $-3$\%; and Ref.~\cite{hollenberg92},
which shows, for example, that the contribution to the $\omega$-meson mass of
the $\omega\to 3\pi$-loop is negligible.  Therefore, ignoring such
contributions does not introduce uncertainty into estimates of the vector
meson mass based on (\ref{bse}).

Equation~(\ref{rhomass}) can also be applied to the $\phi$-meson.  The
transverse component is insensitive to $T$ and $\mu$, and the behaviour of
the longitudinal mass, $M_{\phi+}$, is qualitatively the same as that of the
$\rho$-meson: it increases with $T$ and decreases with $\mu$.  Using $\eta =
1.06\,$GeV, the model yields $M_{\phi\pm} = 1.02\,$GeV for $m_s = 180\,$MeV
at $T=0=\mu$.

In a 2-flavour, free-quark gas at $T=0$ the baryon number density is $\rho_B=
2 \mu^3/(3 \pi^2)\,$, by which gauge nuclear matter density,
$\rho_0=0.16\,$fm$^{-3}$, corresponds to $\mu= \mu_0 := 260\,$MeV$\,=
0.245\,\eta$.  At this chemical potential our model yields
\begin{eqnarray}
\label{mrhoa}
M_{\rho+}(\mu_0) & \approx & 0.75 M_{\rho+}(\mu=0)\,,\\
\label{mphia}
M_{\phi+}(\mu_0) & \approx & 0.85 M_{\phi+}(\mu=0)\,.
\end{eqnarray}
The study of Ref.~\cite{greg97} indicates that a better representation of the
ultraviolet behaviour of the dressed-gluon propagator expands the horizontal
scale in Fig.~\ref{pirhomass}, with the critical chemical potential increased
by 25\%.  Based on this we judge that a more realistic estimate is obtained
by evaluating the mass at $\mu_0^\prime=0.20\,\eta$, which yields
\begin{eqnarray}
\label{mrhob}
M_{\rho+}(\mu_0^\prime) &\approx&  0.85 M_{\rho+}(\mu=0)\,,\\
\label{mphib}
M_{\phi+}(\mu_0^\prime) & \approx & 0.90 M_{\phi+}(\mu=0)\,;
\end{eqnarray}
a small, quantitative modification.  The difference between (\ref{mrhoa}) and
(\ref{mrhob}), and that between (\ref{mphia}) and (\ref{mphib}), is a measure
of the theoretical uncertainty in our estimates in each case.  This reduction
in the vector meson masses is quantitatively consistent with that calculated
in Ref.~\cite{derek95} and conjectured in Ref.~\cite{brown91}.  At the
critical chemical potential for $T=0$, $M_{\rho+} \approx 0.65\,
M_{\rho+}(\mu=0)$ and $M_{\phi+} \approx 0.80\, M_{\phi+}(\mu=0)$.

We have analysed a simple, DSE model of QCD$_T^\mu$ that preserves the
momentum-dependence of gluon and quark dressing, which is an important
qualitative feature of more sophisticated studies.  The simplicity of the
model means that many of its consequences can be demonstrated algebraically.
For example, it elucidates the origin of an anticorrelation, found for a
range of quantities, between their response to increasing $T$ and that to
increasing $\mu$, discovered in contrasting the studies in
Refs.~\cite{greg97} and \cite{bender96}.

We find that both $(-\langle \bar q q)\rangle$ and $f_\pi$ decrease with $T$
and increase with $\mu$, and this ensures that $m_\pi$ is insensitive to
increasing $\mu$ and/or $T$ until very near the edge of the domain of
confinement and DCSB.  The mass of the transverse component of the vector
meson is insensitive to $T$ and $\mu$ while the mass of the longitudinal
component increases with increasing $T$ but decreases with increasing $\mu$.
This behaviour is opposite to that observed for $(-\langle \bar q q)\rangle$
and $f_\pi$, and hence the scaling law conjectured in Ref.~\cite{brown91} is
inconsistent with our calculation, as it is with others of this type.

Our study has two primary limitations.  First, we cannot calculate the width
of the vector mesons in this model because the solution of (\ref{bse}) does
not provide a realistic Bethe-Salpeter amplitude.  We are currently working
to overcome this limitation.  Second, the reliable calculation of
meson-photon observables at $T=0=\mu$ only became possible with the
determination~\cite{ayse97} of the form of the dressed-quark-photon vertex.
The generalisation of this vertex to nonzero $T$ and $\mu$ is a necessary
precursor to the study of these processes at $T\neq 0 \neq \mu$.

\acknowledgements We acknowledge useful conversations with D. Blaschke,
Yu. Kalinovsky and G. Poulis.  For their hospitality and support during
visits in which some of this work was conducted: PM and CDR gratefully
acknowledge the Department of Physics at the University of Rostock; SS, the
Physics Division at Argonne National Laboratory; and CDR and SS, the
Bogoliubov Laboratory for Theoretical Physics and the Laboratory of Computing
Techniques and Automation at the Joint Institute for Nuclear Research.  This
work was supported in part by the US Department of Energy, Nuclear Physics
Division, under contract number W-31-109-ENG-38; the National Science
Foundation under grant no. INT-9603385; Deutscher Akademischer
Austauschdienst; and benefited from the resources of the National Energy
Research Scientific Computing Center.

%______________________________ References ______________________________

%--------------------Figures
%...............
\setcounter{figure}{0}
\begin{figure}
\centering{\
\epsfig{figure=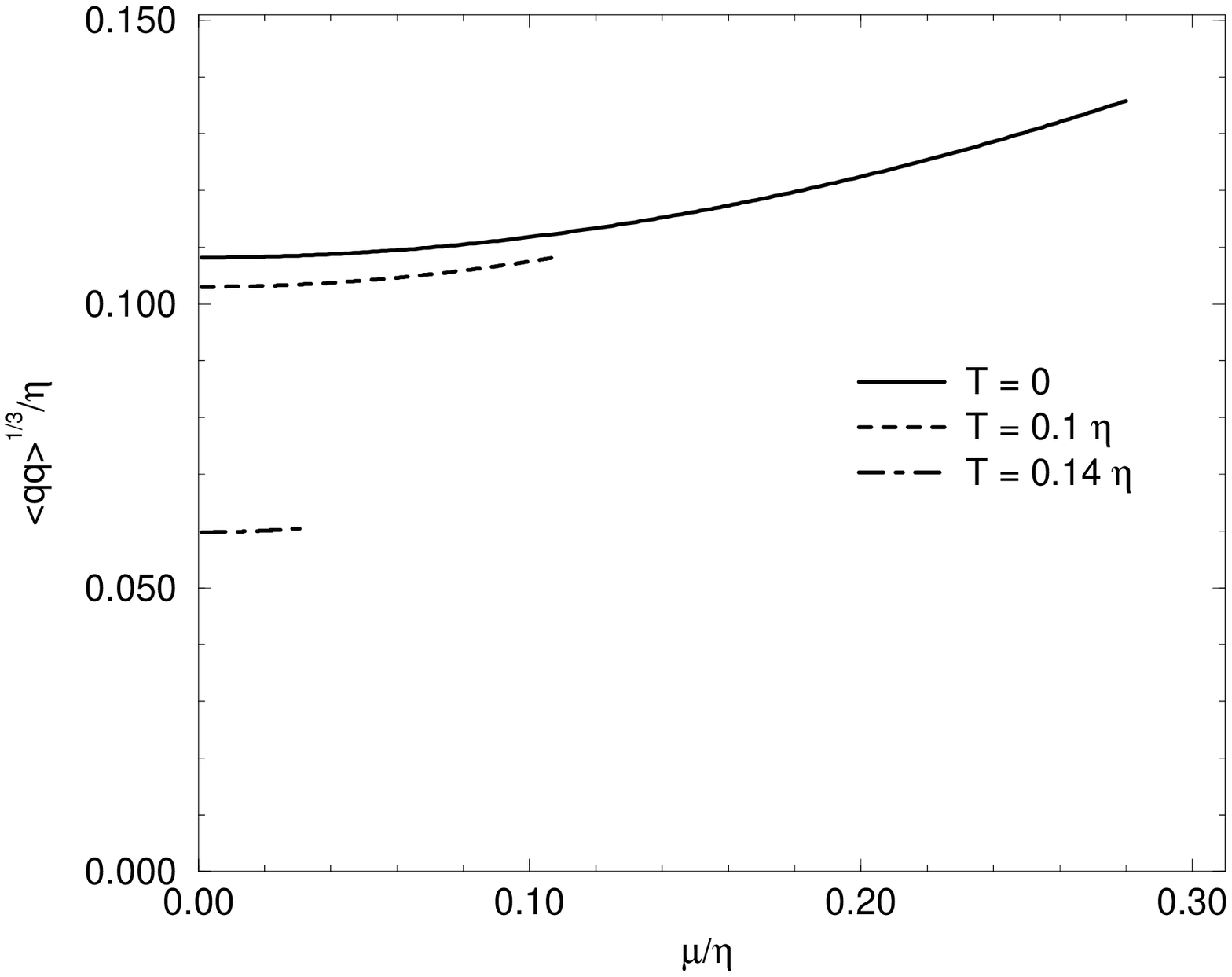,height=13.0cm}}
\caption{\label{condensate} The quark condensate, (\protect\ref{qbq}), as a
function of $\mu$ for a range of values of $T$.  In all existing studies, in
which the quark mass function has a realistic momentum dependence, it
increases with $\mu$ and decreases with $T$.  At the critical chemical
potential, $\mu_c(T)$, $(-\langle \bar q q\rangle)$ drops discontinuously to
zero, as expected of a first-order transition.  For $\mu=0$ it falls
continuously to zero, exhibiting a second-order transition at $T_c(\mu=0)=
0.16\,\eta$.}
\end{figure}
\begin{figure}
\centering{\
\epsfig{figure=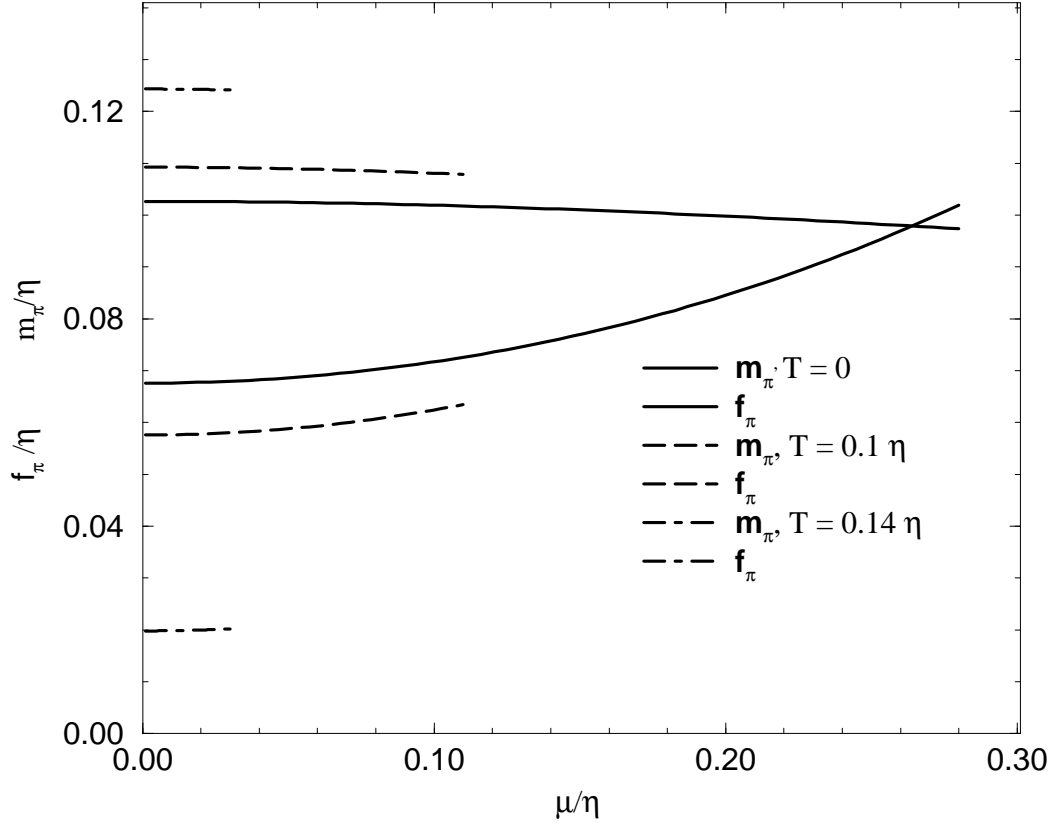,height=13.0cm}}
\caption{\label{fpimpi} The pion mass, (\protect\ref{pimasslhs}), and weak
decay constant, (\protect\ref{npisq}), as a function of $\mu$ for a range of
values of $T$.  $m_\pi$ falls slowly and uniformly with $\mu$
[$m_\pi(T=0,\mu_c)= 0.95 \, m_\pi(T=0,\mu=0)$] but increases with $T$.  Such
a decrease is imperceptible if the ordinate has the range in
Fig.~\protect\ref{pirhomass}.  $f_\pi$ increases with $\mu$ and decreases
with $T$ [$f_\pi(T=0,\mu_c)= 1.51 \, f_\pi(T=0,\mu=0)$].}
\end{figure}
\begin{figure}
\centering{\
\epsfig{figure=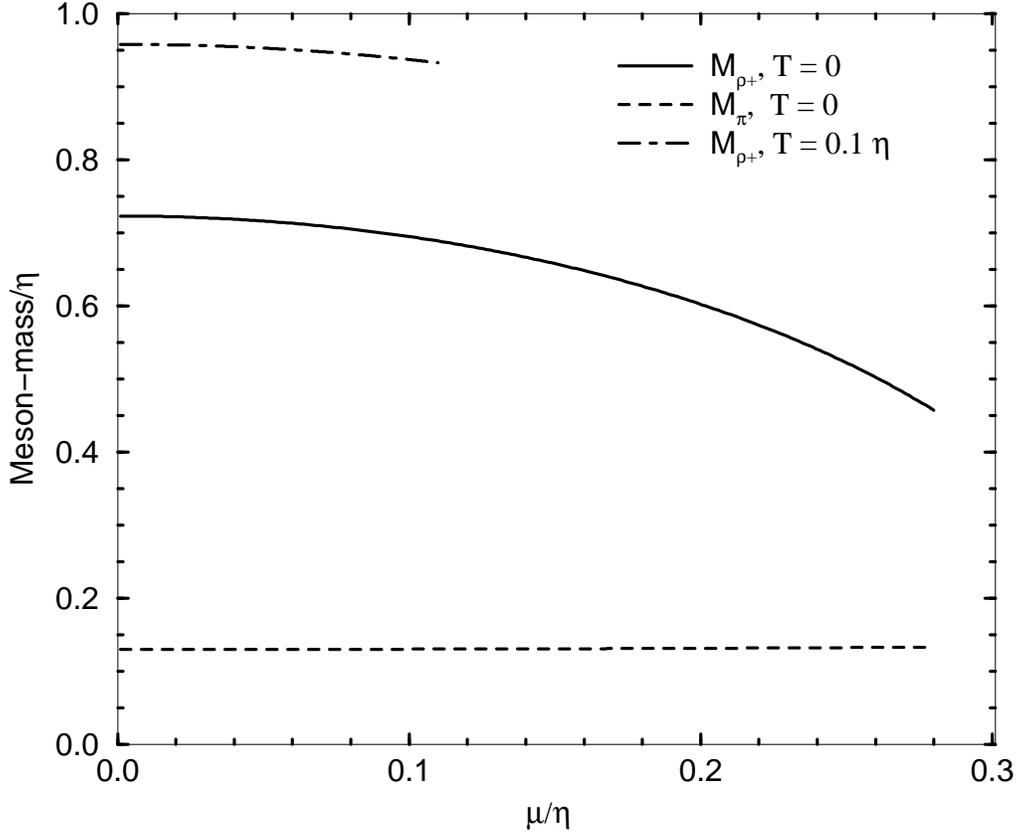,height=13.0cm}}
\caption{\label{pirhomass} $M_{\rho+}$ and $m_\pi$ as a function of $\bar\mu$
for $\bar T = 0, 0.1$.  On the scale of this figure, $m_\pi$ is insensitive
to this variation of $T$.  The current-quark mass is $m= 0.011\,\eta$, which
for $\eta=1.06\,$GeV yields $M_{\rho+}= 770\,$MeV and $m_\pi=140\,$MeV at
$T=0=\mu$.}
\end{figure}
%____________________________________________________________________________
\end{document}